\documentclass{ws-procs9x6}
\usepackage{graphicx}
\newcommand{\be}{\begin{equation}}
\newcommand{\ee}{\end{equation}}
\newcommand{\ba}{\begin{eqnarray}}
\newcommand{\ea}{\end{eqnarray}}

\begin{document}

\title{Standard Model CP and Baryon Number Violation in Cold Electroweak Cosmology
}

\author{Edward Shuryak$^*$ }

\address{Department of Physics and Astronomy, Stony Brook University ,\\
Stony Brook, NY 11794 USA\\
$^*$E-mail:  Edward.Shuryak at stonybrook.edu}

\begin{abstract}
Contrary to popular beliefs, it is possible to explain Baryonic asymmetry
of the Universe inside the Standard Model, provided inflation ended into a
broken phase below the electroweak transition. Two important ingredients
of the solution are multiquanta "Higgs bags", containing W,Z and top quarks,
as well as sphaleron transitions happening inside these bags. Together,
they provide baryon number violation at the level $10^{-2..3}$. Our recent 
calculations show that CP 
violation (due to the usual CKM matrix of quark masses in the 4-th order)
leads to top-antitop population difference in these bags of about
$10^{-9}$. (The numbers mentioned are not yet optimized and simply follow
a choice made by some numerical simulations of the bosonic fields we used
as a reference point.)  
\end{abstract}


\bodymatter

\section{Cold electroweak scenario}
   It is a great honor to give a talk at this inaugural meeting of the Kobayashi-Maskawa Institute.
Moreover, this invitation came at the moment whet I can report new exciting applications
of the celebrated CKM matrix, to its  originally intended purpose -- explaining (perhaps) the most important
CP-odd effect in our Universe, its baryonic asymmetry.

The question how it
was produced is 
among the most difficult open questions of physics and cosmology. The observed effect is usually expressed as the ratio of the baryon
density to that of the photons $n_B/n_\gamma\sim 10^{-10}$. 
Sakharov \cite{Sakharov}
had formulated three famous necessary  
conditions:  the (i) baryon number and (ii) the CP violation, with (iii) obligatory deviations from the stationary ensembles such as the thermal equilibrium.  
Although all of them are present in the Standard Model (SM) and standard Big Bang cosmology,
 the baryon asymmetry which is produced by known CKM matrix is completely insufficient to solve this puzzle.

Significant efforts has been made to solve it using hypothetical  ``beyond the  standard
model" scenarios. In particular, possible large CP violating processes in the neutrino or in the supersymmetric sectors are possible.
Although there are interesting proposals along those lines, they remain orders of magnitude away from feasible experimental tests.
 
  An alternative we will discuss
  is the modification of the standard cosmology. The standard Big Bang scenario predicts adiabatically slow  
crossing of the electroweak phase transition, leading to extremely small deviations from equilibrium.  
The so called ``hybrid" or ``cold" scenario  
\cite{GGKS,KT,Felder:2000hj,GarciaBellido:2002aj}  solve this difficulty by
combining the end of the inflation era with the establishment of
the  electroweak broken phase. Since
there is no space here to discuss it in detail,  let me simply enumerate the main points
of the emerging scenario: only few new points will be elaborated below.

\begin{itemize}
\item
  Large  deviations  from equilibrium, $O(1)$, with strongly oscillating Higgs and other fields \cite{GarciaBellido:2003wd,Tranberg:2003gi,Tranberg:2009de, Tranberg:2010af}. 
\\
\item Formation of relatively long-lived ``hot spots" inside which the Higgs VEV is small \cite{GarciaBellido:2003wd,Tranberg:2003gi,Tranberg:2009de, Tranberg:2010af}. 
 \\
\item  Topologically nontrivial fluctuations of the gauge field lead to baryon number violation \cite{GarciaBellido:2003wd,Tranberg:2003gi,Tranberg:2009de, Tranberg:2010af} at overall level of $O(10^{-3})$
\\
\item These transitions take place only $inside$ the ``hot spots" \cite{GarciaBellido:2003wd,Tranberg:2003gi,Tranberg:2009de, Tranberg:2010af}. 
 \\
\item  The first nonscalar quanta produced are those with large mass (that is, stronger coupled to Higgs), namely the top quarks/antiquarks and the gauge bosons    \cite{Flambaum:2010fp}. \\
\item Tops and gauge bosons are collected into the ``hot spots", which are then mechanically balanced. These objects are identified  \cite{Flambaum:2010fp} with the (non-topological) solitons called the W-Z-top bags  found independently\cite{Crichigno:2010ky}\\
\item Baryon number violation events are identified with Carter-Ostrovsky-Shuryak (COS) sphalerons with the size tuned to those of the hot spots  \\
\item The B violation processes, described by the well known 12-fermion 't Hooft operator, can occure as various subprocesses, $n\rightarrow 12-n, n=0..12$ . The most probable was argued  \cite{Flambaum:2010fp} to be the ``top recycling  $\bar{t}\bar{t}\bar{t} \rightarrow 9$  which is converting the energy of the tops or antitops already present in the bag
into that of the gauge field, which helps passing high barrier separating different topologies\\
\item If so, the asymmetry of top and anti-top populations in the bag  leads to different production rate of the baryon and antibaryon numbers \cite{arXiv:1107.4060}  \\  
\item Clear ``time arrow" of the process is given by the fact that a top quark, after a weak decay into lighter quarks, simply diffuse away from the bag since all quarks lighter than the top cannot be bound to the bag  \cite{arXiv:1107.4060}
 \\  \item  The usual CKM mechanism of the CP violation naturally produces the top-antitop asymmetry via interferences of various
outgoing paths of these light quarks, in the 4-th order in weak interaction \cite{arXiv:1107.4060} \\
\end{itemize}



\section{The multi-quark bags}

%
 Being a scalar, the Higgs generates {\em universal attraction} between all kinds of particles. Furthermore, the strength of the attraction is proportional to their total mass, similar to the gravity interacting with the total energy.
Gravity, feeble as it is, holds together planets, stars and even create black holes.  Unlike vector forces induced by electric, weak  or color charges,  gravity and scalar exchanges are exempt from  ``screening" and thus their  weak coupling can be compensated by a large
number $N$ of participating particles.   And yet, unlike gravity, the Higgs boson is neither massless, nor even particularly light in comparison to $W,Z$ or $t$. So, are there  multi-quanta states based on the Higgs attraction?

Another instructive analogy is provided by the nuclear physics. Think of a 
  (much-simplified) Walecka model, in which the nuclear forces can be approximately described by the $\sigma$,  ``the Higgs boson of the nuclear physics", and $\omega$ meson exchanges.
  Because of similarity of masses $m_\sigma\sim 600\, MeV,m_\omega\sim 770\, MeV$, as well as couplings, the
  sigma-induced attraction  is {\em nearly exactly canceled} by the omega-induced repulsion. Their sum is an
   order of magnitude smaller than one would get from scalar and vector components taken separately.
   
   Can the situation at electroweak scale similar? Perhaps the Standard Model is just a
low energy effective Lagrangian, hiding some deeper physics behind
its simplistic scalar Higgs.  We are not aware of any particular model which suggests a vector companion to Higgs
with a similarly small mass $O(100\, GeV)$ mass.   For example,  the ``techni-$\rho$" is predicted to be at the scale
$M_\rho\sim1\, TeV$. Thus, unlike in the nuclear physics,
 one is $not$ expecting the scalar-vector cancellation. 

  The interest in the
issue of ``top bags" originated from the question whether a sufficiently heavy
SM-type fermion should actually exist as a bag  state,
depleting the Higgs VEV around itself. 
  Although classically this seemed to be possible, it was 
shown in refs \cite{Dimopoulos:1990at, Bagger:1991pg,Farhi:2003iu} 
that quantum (one loop)
effects destabilize such bags, except at so large coupling at which the
Yukawa theory itself becomes apparently sick, with an instability of its ground state.
The issue rest dormant for some time till Nielsen and Froggatt \cite{Froggatt:2008ns} suggested to look at the first magic number, 12
tops+antitops  corresponding to the maximal occupancy of the lowest $
l=0,\, j=1/2$ orbital, with 3 colors and 2 from $t+\bar{t}$. Using simple
formulae from atomic physics these authors suggested that such system
 forms a deeply-bound state. 
  In ref.\cite{Kuchiev:2008fd} we have checked this claim and found that, unfortunately, this is $not$ the
case. While for a massless Higgs there are indeed weakly bound states of 12 tops, 
 they disappear way below the realistic Higgs mass.
 Further  variational improvement of
the binding conditions for the 12-quark system 
\cite{Richard:2008uq} confirmed that 12 tops are unbound for Higgs mass 
$m_H>M_c(12)\sim  50\, GeV$. 


Assuming spherical symmetry, the Higgs energy reads
\begin{equation}
E_{Higgs}=2 \pi v^2\int_0^\infty dr \, r^2\left[\phi^{\prime \,2}+\frac{1}{4}%
\,m_H^2\,(\phi^2-1)^2 \right] 
\label{Higgs_Hamiltonian}
\end{equation}
where  $v=246\,GeV$ ,$m_H^2 \equiv 2 \lambda v^2$ is the Higgs mass, taken to be a round number\footnote{If the ATLAS/CMS peak in diphoton will become a real HIggs mass, then it is about 125 GeV.} $m_H=100\,GeV$
in the original papers (we also use units of $100\,GeV$ throughout this paper). 
Consider now the addition of a conserved (during the time scale we are interested in) $N$ particles (fermions or bosons), couple strongly to the Higgs field which  could be strongly distorted.  We adopt a mean-field approximation, in which all the particles are described by the \textit{same} wave functions in the background of the Higgs field. Corrections to this mean-field description, such as, many-body, recoil and retardation of the Higgs field are expected to be suppressed by factors $v/m$, $m_H/m$ and $1/N$. 
In the semiclassical approximation, the total energy of the system will  be given by 
$
E_{cl}=E_{Higgs} + \sum_{a}n_{a}\varepsilon _{a},
$
where $\{\varepsilon_a\}$ is the spectrum of the corresponding field in the Higgs background,  $n_{a}$ is the occupation number of each state and $N=\sum_{a}n_{a}$ is the total, conserved, particle number.

In the Higgs vacuum, i.e. $\phi(r)=1$, the state of lowest energy with $N$ particles has total energy $NM$. However, in the background of a non-trivial Higgs field there are two competing effects. On the one hand,
 the gradient and potential terms increase the energy but, on the other hand, there might be some bound states levels with energy $0 < \varepsilon_{a} < M$ which can allocate the quanta, lowering the energy of the system of particles at the expense of creating such distortion. 
 
Let us start by a crude estimate of the the order of magnitude of $N$ for which such bags may exist. If we were to deplete  a certain large volume of the Higgs VEV (surface/kinetic terms neglected for now), it would require an energy
$V_{bag}   \frac{m_H^2}{8} v^2$.
For a bag of radius, say, $R\cdot 100\, GeV$=4, this energy is about 20 $TeV$. Thus, if the lowest  W-boson energy level has a binding energy of the order of $30\,GeV$ per $W$ or $Z$, an order of $O(1000)$ of them would be needed to compensate for the bag energy and
obtain some binding. The top quarks are  heavier and may get much larger binding, so one might naively think that less of them would suffice: but Pauli exclusion principle  makes it  more delicate.

Consider the propagation of $W$-bosons in an external Higgs field 

%
   \begin{eqnarray}
      \label{form}
      \left( \Box+M_W^2\phi^2\right) W^\mu
     +\partial^\mu \left(\frac{ W^\nu\partial_\nu \phi^2}{\phi^2}\right) =0.~~\quad
    \end{eqnarray}
 Let us study these equations of motion in the usual electric (e), longitudinal (l)  and magnetic (m)  basis
$
\boldsymbol{W}^{(e,l,m)}  = \boldsymbol{ Y}^{(e,l,m)}_{jm} f_{(e,l,m)}(r)/r, 
$
where $\boldsymbol{ Y}_{jm} $ are spherical harmonic vectors and $ f_e(r), f_l(r)$ and $f_m(r)$ are the radial wave functions for each mode. In a static, spherically symmetric, background the last term in (\ref{form}) vanishes for the magnetic mode, leading to the simple Klein-Gordon equation: but in this case
   $j\geq1$.  For others one can start with smaller $j=0$ without a centrifugal potential, but the Laplacian mixes the electro-longitudinal modes, leading to the set of coupled equations (see the original paper)
%
%
in which the last term in (\ref{form})  becomes large and positive in the region where the Higgs field  $\phi$ approaches zero, effectively repelling the longitudinal modes from the  bag. (Note that massless gauge fields have no longitudinal degree of freedom at all.) 
As a consequence, even $j=0$ mode is pushed above that $j=1$  magnetic modes, which is thus the lowest. 
In order to find bag solutions for finite $N$, we adopted a variational approach and took as  a trial function for the Higgs, e.g. the Gaussian profile
 \be
  \phi(r)=1-\alpha \, \exp{(-r^2/w^2)},  \label{eqn_Gaussian} 
  \ee 
with two parameters, $\alpha$ and $w$ describing its depth and the width, respectively. Solving the W-boson magnetic equation in this Higgs background is rather straightforward 

\begin{figure}[t!]
\includegraphics[width=5.cm]{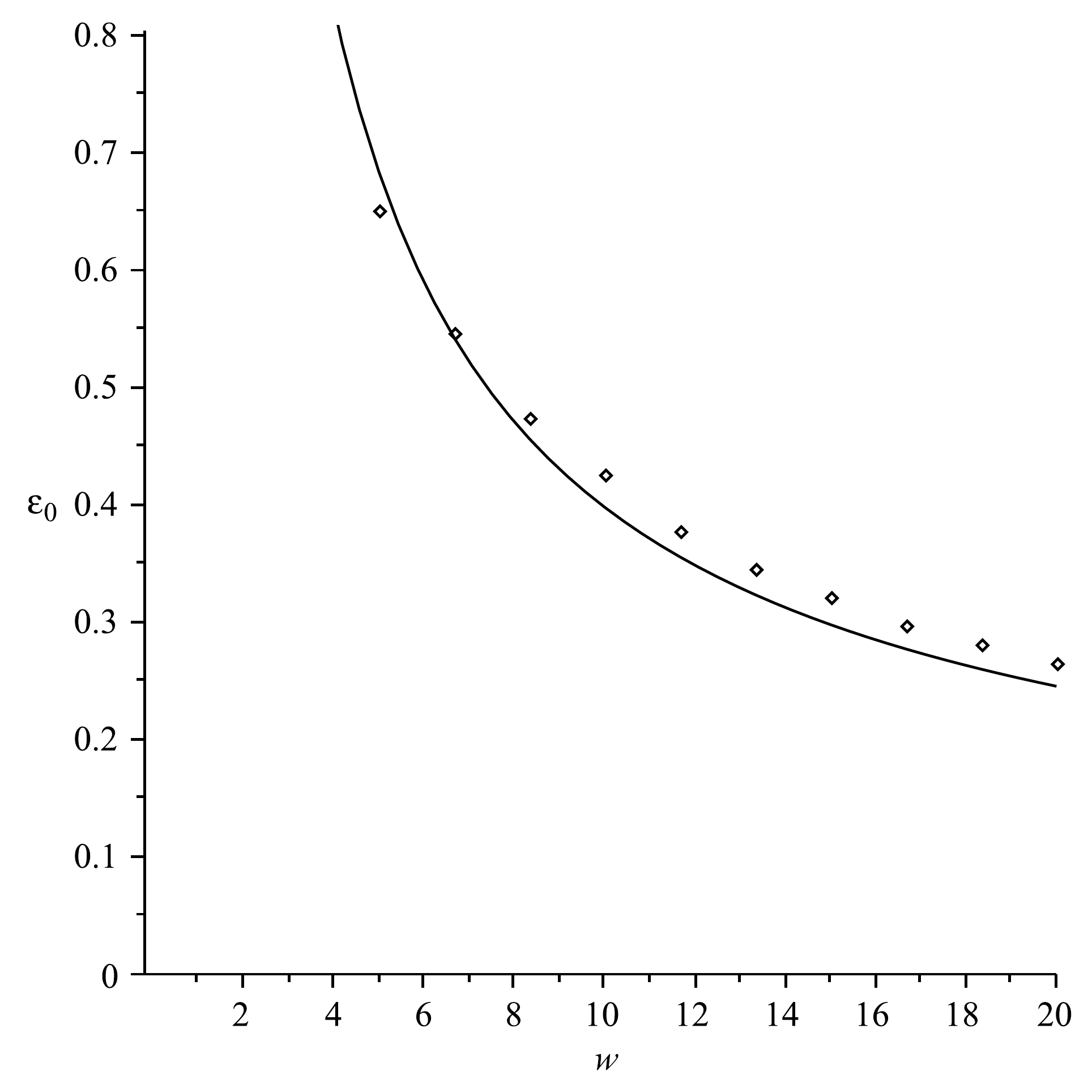}
\includegraphics[width=5.cm]{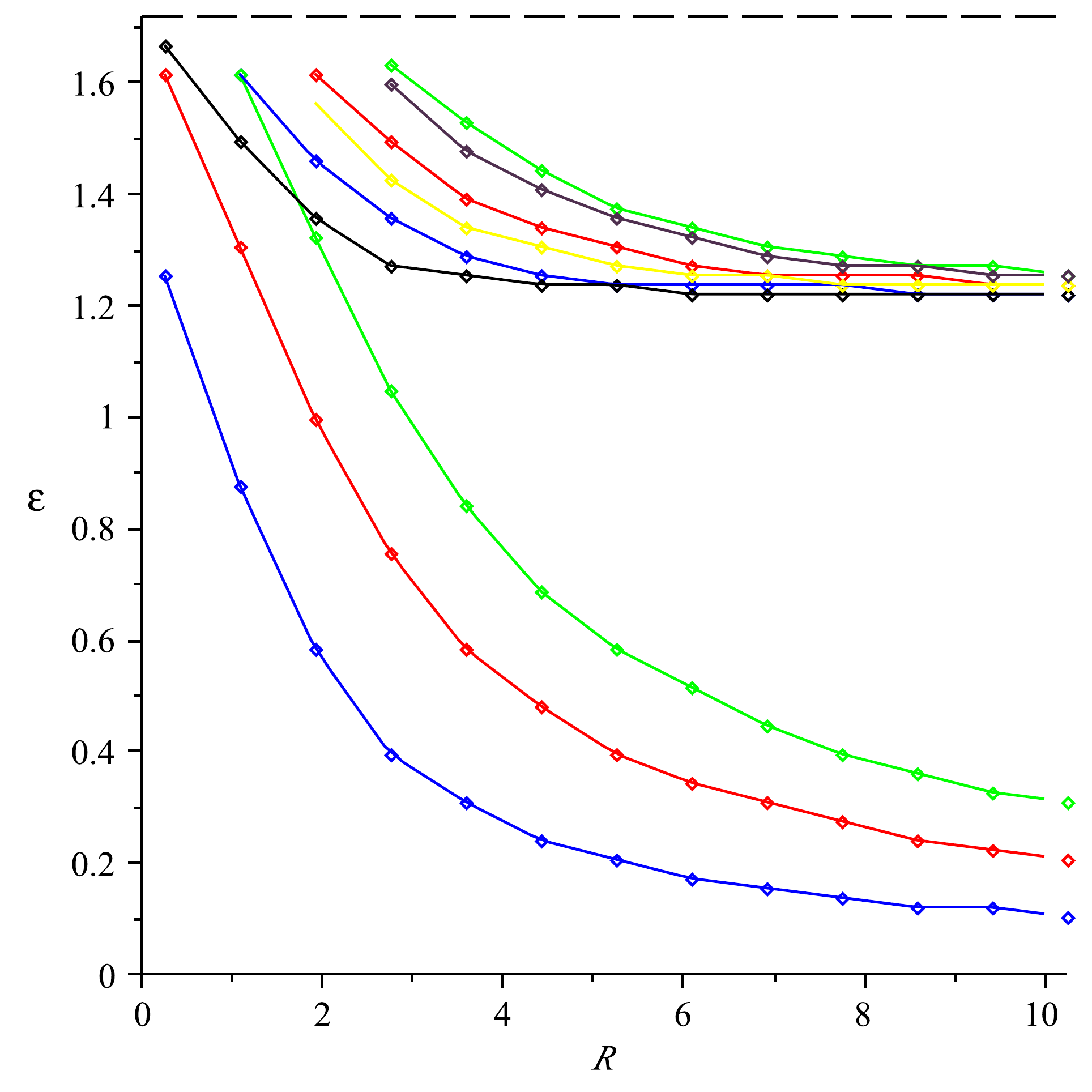}
\caption{(a) The energy of the lowest magnetic mode ($j=1$) $E/(100\, GeV$  for a Gaussian bag with $\alpha=1.3$ versus its width. The numerical results (points) are in good accordance with the analytical result, see \cite{Crichigno:2010ky}. (b)
 Dependence of fermionic bound state levels energy $\varepsilon$ in units of $100\,GeV$ on the size of the bag, expressed as the parameter $R$ in units $1/100\, GeV^{-1}$ for a bag with $\eta=1$. The color coding of the levels and their quantum numbers are listed in Table I.}
\label{fig:Top_Levels_Kink}
\end{figure}

In Fig. \ref{fig:Top_Levels_Kink}(a) we show some results for a bag with $\alpha=1.3$. It is now relatively simple thing to vary the shape and reach the minimum of the energy of the system
 (still in the spherically symmetric ansatz.)


Now we consider  a system of $N$ heavy fermions interacting with a background Higgs field, 
the standard notations for Dirac
spinors in spherical coordinates are
\begin{equation}
\psi =\frac{1}{r}\left( 
\begin{array}{c}
F(r)\Omega _{jlm} \\ 
(-1)^{1/2(1+l-l^{\prime })}G(r)\Omega _{jl^{\prime }m}%
\end{array}%
\right),
\end{equation}%
where $\Omega _{jlm}$ are spherical 2-component spinors and we take normalization $\int dr \,(F^{2}+G^{2})=1$. The so-called Dirac parameter $\kappa$ is defined as 
\begin{equation}
\kappa =\left\{ 
\begin{array}{c}
-(l+1)\text{ \ \ \ for }j=l+1/2 \\ 
l\text{ \ \ \ \ \ \ \ \ \ \ \ \ \ \ for\ }j=l-1/2%
\end{array}%
\right. 
\end{equation}
and runs over all nonzero integers, being positive for anti-parallel spin and negative for parallel spin. The Dirac's equation reads 
\begin{eqnarray}
 (\varepsilon-m\,\phi)\,F~\, & = &\,-G^{\prime}\,+\,(\kappa/r)\,G \\ \nonumber
(\varepsilon+m\,\phi)\,G~\,& = & \,\,~~F^{\prime}\,+\,(\kappa/r)\ F  
\label{Dirac_Equation}
\end{eqnarray}
The form of these equations presumes that the eigenvalue $\varepsilon$ is positive. A negative eigenvalue would correspond to a state in the lower fermion continuum.  If so, a charge conjugation transformation turns it into a positive eigenvalue for an antifermion. The Higgs equation of motion reads
 \begin{equation}
\phi^{\prime\prime}+\frac{2}{r}\phi^{\prime}+\frac{m_H^2}{2}\phi\,(1-\phi^2)\,=\,\frac{(N-1)m}{4\pi \,v^2}\,\frac{F^2-G^2}{r^2}
\label{Higgs_equation_motion} 
\end{equation}
Note that in the $r \rightarrow \infty$ limit, the source term in the right hand side, as well as the additional $\phi^{\prime}/r$ term from the Laplacian can be neglected and the equation becomes the usual equation for a 1D kink. 

 The spectrum for Dirac's equation is found numerically  and examples of the levels
are shown in Fig.\ref{fig:Top_Levels_Kink}. The Table shows magic numbers and the order in which levels are populated. Some levels  are also shown in
  Fig. \ref{fig:Top_Levels_Kink}.

\begin{table}[h!]
\begin{tabular}{r|llllcl|c}
& $n_r$ & $\kappa$ & $l$ & $j$ & $Deg. (\bar t t)$ & color &  \\ \hline
1 & 0 & -1 & 0 & 1/2 & 12 & blue &  \\ 
2 & 0 & -2 & 1 & 3/2 & 24 & red &  \\ 
3 & 0 & -3 & 2 & 5/2 & 36 & green &  \\ 
4 & 0 & 1 & 1 & 1/2 & 12 & black &  
\end{tabular}%
\caption{The properties of some levels, including the number of
radial nodes $n_r$, Dirac parameter $\protect\kappa$, orbital momentum $l$,
total angular momentum $j$, multiplicity of states $Deg$(for $t \bar t$ bags) and color code used in
our figures. }
\end{table}

Attempting to find a minimum of the total energy, tops and Higgs, we have found that
the ratio of the top-to-Hoggs masses are simply not large enough to stabilize  bags by themselves.  So tops can only exist
 inside  the W-bags.
Free (or weakly bound) top quarks are much heavier than $W$ bosons and thus decay
into another quark and the $W$. In the Higgs bag, however, we found that two lowest  top levels are below the
  lowest  gauge boson ones; so up to 36  $t+\bar{t}$ in such bags may live much longer, given by
 next order decays into three fermions, like in the usual beta decays.

\section{The sphalerons and rates}
In the broken phase the electroweak sphalerons had been found by Manton et al:  their mass
is about 14 TeV  and thus the tunneling rate is prohibitively small. In pure gauge sector finding
the sphaleron solution has been precluded by the fact that classical gauge theory has no scale
and thus energy has no minimum. It has been surpassed in \cite{Ostrovsky:2002cg} by requiring minimization
under two conditions: fixed Chern-Symons number and mean-square radius $\rho$.

The profile of the the magnetic field in  COS configuration is given by the following spherically symmetric expression
 \be B^2(r)= {48 \rho^4  \over g^2(r^2+\rho^2)^4} 
\label{eqn_profile}\ee
 This form does indeed
fit well the numerically found shapes of the $B^2$ at the ``sphaleron moment",
a maximum of magnetic field.
Fitted  radius,
yielding $m \rho\approx 3.9$ which corresponds to
the total energy of the COS sphaleron 
\be E_{tot}=3\pi^2/g^2 \rho \approx 2\, \,TeV \ee
which is 7 times less than the KM sphaleron mass:  it makes a huge difference for he rate.



Combining all the factors we find that numerical value of preexponent and exponent
nearly cancels out, leaving crudely \be \Gamma / V m^4\sim 10^{-1} \ee with accuracy
say an order of magnitude or so. With that accuracy it agrees with
 the results of the simulations which
 also finds that the number of sphaleron transitions per spot  is indeed  about several percents.

   What should happen after the sphaleron moment?
 Sphaleron decay  is a classical downhill rolling of the classical (high amplitude) gauge field, from the (sphaleron) top into
 the next classical vacuum . This process was extensively studied numerically for the broken-phase
 sphaleron. Remarkably, an $analytic$ solution of the time-dependent  explosion
of  COS sphaleron has also been found in  the COS paper \cite{Ostrovsky:2002cg}.
 The late time profile of the energy density of the expanding ``empty" shell. 
\be 4\pi \epsilon(r,t)={8\pi \over g^2 \rho^2 r^2} \left({1\over
  1+(r-t)^2/\rho^2 }\right)^3  \label{eqn_shell}
\ee
 Comparing the explosion of COS sphaleron with numerical data
  one can see both the similarities and the differences between them.
   Indeed, there is an empty shell formation at some time. However
  the inside of the shell does not remain empty: in fact the topology and magnetic field
 have a secondary peak (of smaller magnitude). Qualitatively it is easy to see why it happens.  
The COS sphaleron is a solution exploding
 in zero Higgs background, with massless gauge fields at infinity.
 In the numerical simulations we discuss  such explosion happens inside
  the  finite-size cavity. As the gauge bosons of the expanding shell hit the
  walls of the no-Higgs spot, they are massive outside.  With some probability they


Since the original numerical simulations have included the gauge fields but ignored fermions, 
we have to discuss first, at quite qualitative level, what their effects can be,
relative to that of the gauge fields.  Those tops  would be added to 
 the  metastable bubbles of the symmetric phase, the no-Higgs spots,  like  the W discussed 
 above.

  The well known Adler-Bell-Jackiw anomaly  require
 that  a change in  gauge field 
topology  by $\Delta Q\pm 1$ must be accompanied by a corresponding
change in baryon and lepton numbers, B and L. More specifically, 
such topologically nontrivial fluctuation
 can thus be viewed as a ``t'Hooft operator" 
with 12 fermionic legs. Particular
fermions depend on orientation of the gauge fields in the electroweak SU(2): since we are interested in
utilization of top quarks, we will assume  it to be ``up". In such case the 
produced set contains $t_rt_bt_g c_rc_bc_g  u_ru_bu_g \tau\mu,e$ , where $r,b,g$ are quark colors.
to which we refer below as the  $0\rightarrow 12$ reaction.
Of course, in matter with a nonzero fermion
 density  many more reactions
of the type $n\rightarrow (12-n)$  are allowed, with $n$ (anti)fermions captured from
the initial state.



 The 
classical solution describing the expansion stage at  $t>0$ has been worked out for COS sphaleron explosion,
and for the ``compression stage'' at $t<0$ one can use the same solution with a time reversed. 
  At very early time or very late times
 $t\rightarrow \pm\infty$  the classical field become weak and
describe convergent/divergent spherical waves, which are nothing but
certain number of colliding gauge bosons. 
 Fermions of the  theory should also be treated accordingly.
  Large  semiclassical parameter -- sphaleron energy over temperature -- parametrically leads to the
 assumption that total bosonic energy is much larger than
that of the fermions, so one usually ignores backreaction and
consider Dirac eqn for fermions in a given gauge background. For KM sphaleron and effective $T$ we discuss,
this parameter would be $\sim 70$, which is indeed large compared to 12 fermions. However in the case
of COS sphaleron we are going to use the number is about $\sim 10$,  comparable to the number of fermions produced.
It implies that backreaction from fermions to bosons is very important.

The only
(analytic) solution to 
Dirac eqn  of the ``expansion stage'' 
was obtained in \cite{Shuryak:2002qz}, it describes motion
 from the COS sphaleron zero mode (at $t=0$) 
all the way to large $t\rightarrow+\infty$ physical 
outgoing fermions, with analytically calculated momentum distribution. A new element pointed out in \cite{Flambaum:2010fp}
is  that its
time-reflection  can
 also describe the ``compression stage", in which
 free fermions with the $negative$ energy are captured by a  convergent
 spherical wave of gauge field at  $t\rightarrow-\infty$,
  ending at the zero energy sphaleron zero mode at $t=0$. 
%


  This  implies that energy
of the initial fermions can be incorporated and used
in the sphaleron transition.  The optimum way
to generate sphaleron transition turns out to be 3 initial top quarks\footnote{In order to satisfy Pauli
principle and fit into the same sphaleron zero mode, colors of the 3
quarks of each flavor should all be different. }
considering the $3\rightarrow 9$ fermion process instead of 
the original  $0\rightarrow 12$ one.
 The $3\rightarrow 9$ fermion process saves a lot of energy,  
as in it  the initial top quark energy can be completely
 transferred 
from the ``sphaleron doorway state'' to the gauge field. Estimates show that
it increases the sphaleron rate by about one order of magnitude, compared to pure gauge calculation.

%


\section{The CP asymmetry}

The first attempts to estimate  magnitude of CP violation
 in cold electroweak cosmology has been made by Smit, Tranberg and collaborators \cite{Hernandez:2008db, Tranberg:2009de}. 
Their strategy has been to derive some local effective CP-odd Lagrangian by integrating out quarks,
and than include this Lagrangian in their real-time bosonic numerical simulations. The estimated magnitude of the CP-odd effects ranges from $n_B/n_\gamma \sim 10^{-6}.. 10^{-10}$ \cite{Tranberg:2010af}, which reignites hopes that this scenario can provide the observed magnitude of the baryon asymmetry in Universe.
However,  there are many unanswered questions about the accuracy of these estimates. One of them \cite{Flambaum:2010fp} is that the  effective Lagrangian  derived with specific scale of the loop momenta, e.g. $p\sim m_c$, can only be used
for field configurations at a scale softer than this loop scale: and the ``hot spots" in numerical simulations obviously
do not fit this condition.
But in practice even more important is the following unanswered generic question:  why should a very complicated operator (containing 4-epsilon symbol convoluted with 4 gauge field potentials and one field strength) averaged over very complicated field configurations (obtained only numerically) have  $nonzero$ average at all? We were thinking about some
model fields (e.g. sphalerons) in which these operators have nonzero values, but were not able to find any convincing examples. 
Since the calculation is numerical, it would be desirable to have some parametric estimate of the effect, in particular to know
what sign the effect should have and at least some bound on it from below.  These goals are reached in the last paper \cite{arXiv:1107.4060}.


For a top quark starting  at the position $r_1$, the escaping amplitude has  the form
\be
A_t(r_1)=\int \gamma^\nu W^-_\nu V^\dagger P_t S_d^L(r_1,r_2) \gamma^\mu W^+_\mu V  S_u^L(r_2,r_c)d^4r_2,
\ee
whereas the anti-top has a C-reflected expression
\be
A_{\bar t}(r_1)=\int\gamma^\nu W^+_\nu \bar{V^\dagger}P_{\bar t} S_d^L(r_1,r_2) \gamma^\mu W^-_\mu \bar{V}  S_u^L(r_2,r_c)d^4r_2.
\ee
Here $V$ is the CKM matrix, $S$ the quark propagators, their index $u,d$ etc denotes the up and/or down quark flavors and $P_t$ denote the 
flavor matrix making the initial projection on the top quark. 

\begin{figure}[t]
\includegraphics[width=6.cm]{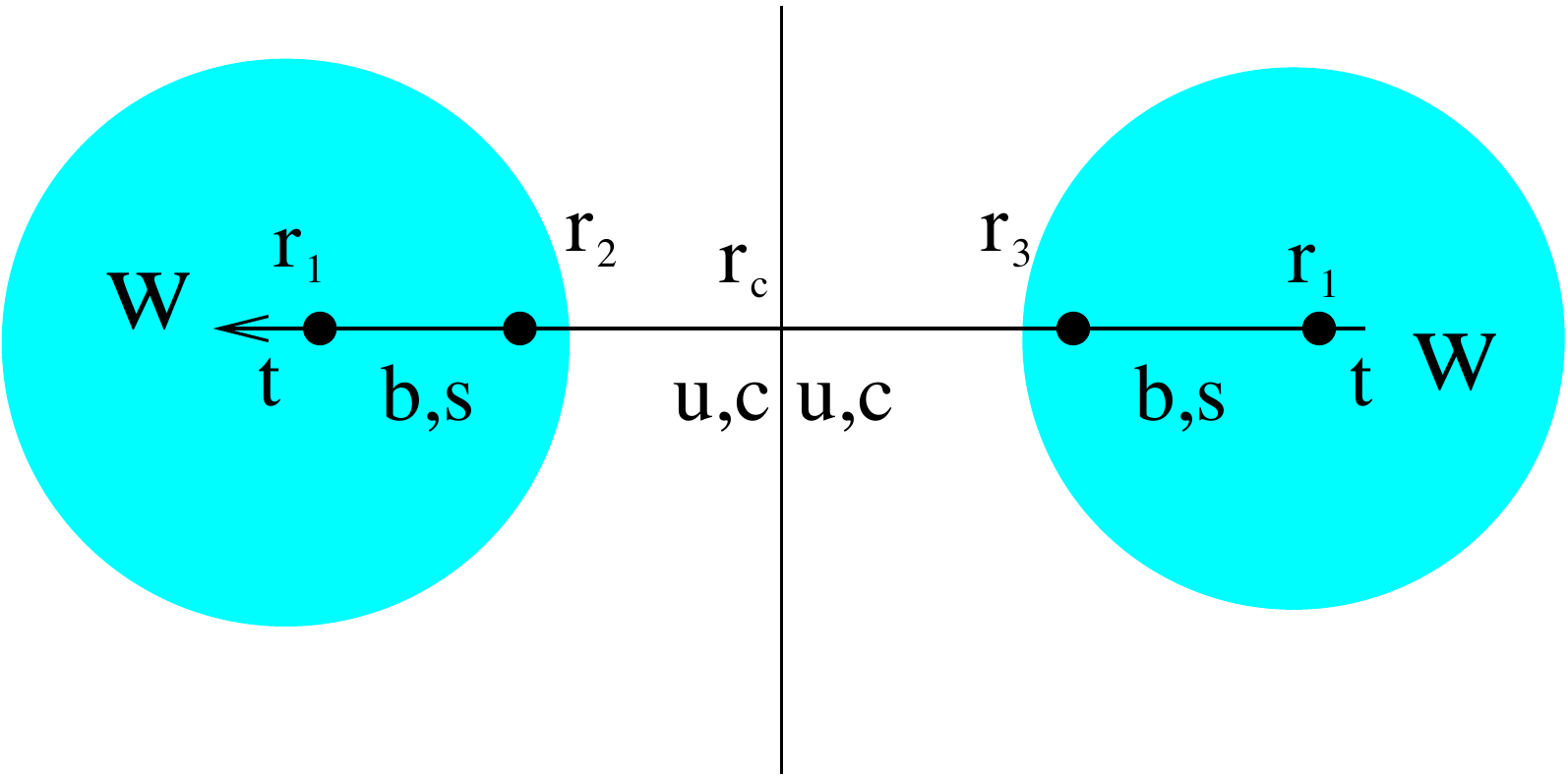}
\caption{(Color online) Schematic shape of the 4 order process involving only quarks of the 2nd and 3ed generations. The shaded objects on the left  and right represent the Higgs bag with strong gauge fields (indicated by W in the figure) inside. The vertical line is the unitarity cut. Four black dots indicate 4 points, indicated with $r_i,i=1..4$ where $W$ quanta are interacting with the quark,
changing it from upper ($t$=top) to lower ($b,s$) components.} 
\label{fig_2bags}
\end{figure}

The probablility of a top quark escaping from $r_1$ is then given by the integral over all positions and sum over all intermediate and
final states $f$ of the squared amplitude
\ba
Prob_t(r_1)&=&\int_{f} A_t^\dagger A_t = Tr \int d^4r_c \sum_{u} A_t^\dagger A_t 
=\int d^4r_c d^4r_2 d^4r_3 Tr\left[P_t\gamma^\nu W^-_\nu V^\dagger S_d^L(r_1,r_2)\notag
\right.
\\&& \gamma^\mu W^+_\mu V  S_u^L(r_2,r_c) S_u^{L\dagger}(r_c,r_3)
\left. \gamma^\alpha W^-_\alpha V^\dagger S_d^{L\dagger}(r_3,r_1)\gamma^\beta W^+_\beta V\right].\notag
\ea
Note that the interference terms between different paths are of the 4-th order in weak interaction, and thus have 4 CKM matrices,
as indeed is needed for the CP violation effects.
Four positions of the points at which the interactions take place, as well as particular quark flavor in the intermediate line, are summed over. Writing the amplitude squared
of the process, one includes the unitarity cut (the vertical line in Fig.\ref{fig_2bags}) to the right of which one, as usual, finds the conjugated image
of the process in opposite direction. 
  
  In between these four points the flavor of the quark remains unchanged. Quark wave functions (we keep in mind $l=0$ or $s$-wave ones only, thus
  points are only indicated by their radial distance from the bag)  are different for each flavor, because of different Yukawa couplings to Higgs profile.
  Semiclassically the phase is approximated 
  \be 
  S_{12}=exp[i \int_{r_1}^{r_2}p(x)dx]\approx exp[i \int_{r_1}^{r_2}(E-{\frac{m_i^2(x) }{ 2E}}) dx],
  \ee
  where $E$ is the quark energy, and the approximation implies that all lower quark flavors are  light in respect to $r/E$, so that  the flavor-dependent phase
  (stemming from the second term in the bracket) is still smaller than 1 

  Let us follow the flavor part of the amplitude, which distinguishes between quarks and anti-quarks. 
  The 4-th order process we outlined in the preceeding section corresponds to the trace of the following matrix product
  \be 
  M_t=Tr(P_t* V*S_{12}^{down}*V^+*S^{up}_{23}*V*S^{down}_{31}*V^+),
  \ee
  where $S$ are quark propagators, the lower indices specify their initial and final points,  the upper subscript remind us
  that those are for up or down quark components. $P_t$ is the projector requiring that we start (and end the loop) in the bag, with a top quark.
  We also define the amplitude for the top-antiquarks
  \be 
  M_{\bar{t}}=Tr(P_t* V^{*}*S_{12}^{down}*V^{T}*S^{up}_{23}*V^{*}*S^{down}_{31}*V^{T})
  \ee
which we  subtract from $M_t$, as 
the effect we evaluate is the difference in the top-antitop population inside the bag.   
The difference gets CP-odd as seen from its dependence on the CP-odd phase $\delta$
  \ba
  M_t- M_{\bar{t}}= 2 i J (S^u_{23}-S^c_{23}) 
 (-S^s_{31} *S^b_{12}-S^d_{31} *S^s_{12}+S^d_{31} *S^b_{12}   \nonumber \\ 
 +S^d_{12} *S^s_{31}-S^d_{12} *S^b_{31}+S^s_{12} *S^b_{31})
 \label{eqn_Mt-Mantit}
   \ea
   \ba 
J=  \sin(\delta) \sin(\theta_{12})\sin(\theta_{13})\sin(\theta_{23}\cos(\theta_{23}) 
 \cos(\theta_{12})\cos^2(\theta_{13})\cos(\theta_{23}) \nonumber
  \ea

The remaining combination of propagators, organized in two brackets, needs to be studied fiurther. Note first that the propagators in the range 2-3 (through the
unitarity cut) factor out and that one may ignore the top quarks there. Note further, that if the $u,c$ quarks would have the same mass, 
the first bracket would vanish: this is in agreement with general arguments that any degenerate quarks should always nullify the CP-odd effects,
as the CP odd phase can be rotated away already in the CKM matrix itself.

The last bracket in (\ref{eqn_Mt-Mantit}) contains interferences of different down quark species: note that there are 6 terms, 3 with plus and 3 with minus.
 Each propagator, as already noticed in the preceeding section, has only small corrections coming from the quark masses. Large terms which are flavor-independent
  always cancel out, in both brackets in the expression above.   Let us look at only the terms which contain the heaviest $b$ quark in the last bracket,
  using the propagators in the form 
$ S^q_{ij} \sim  exp\left( \pm i \frac{m_b^2}{2 E} r_{ij} \right)$
  where $\pm$ refers to different signs in the amplitude and conjugated amplitude and $r_{ij}=r_j-r_i$. Note that the sign of the phase between points $r_2$ and $r_3$ can be positive or negative as it results from a subtraction of the positive phase from $r_3$ to the cut $r_c$ with the negative phase form the cut $r_c$ to $r_2$. Terms containing odd powers  in $r_{23}$ therefore should vanish in the integral: and the lowest term we have is quadratic.
  Considering all phases  to be small due to $1/E$  and using the mass hierarchy $m_b \gg m_s \gg m_d$ 
  we pick up the leading contribution of the last bracket in (\ref{eqn_Mt-Mantit}) which has $r_{23}^2$
  and the 4-th power in the last bracket,  the 6-th order in the phase shift in total:
  \be 
   M_t- M_{\bar{t}}\propto  J   \frac{ m_b^4 m_c^4 m_s^2 r_{23}^2 r_{12}  r_{31} (r_{12}+   r_{31})}{ 64 E^5}.
\label{F}
  \ee 
 Note that all distances in this expression are defined to be positive and the sign in the last bracket is plus, so unlike all the previous orders in the phase expansion,
 at this order we have sign-definite answer with no more cancellations possible. This point is the central one in this work.
We further see that this expression grows for large $r$'s, which are to be integrated over. Of course as we expanded the exponent in the phases, they have to be such that these phases are smaller than 1. 


Let us start with a ``naive" estimate, which assumes that $E$ in the formulae is given by the
top quark mass $ E\sim m_t=173\, GeV $.
As for the field strength, naively one may take all four interaction points inside the bags, where
the amplitude of the $W$ is the strongest. If so, all distances $r_{ij}$ are of the order of the bag size $R_{bag}\sim 1/m_W$.
However, if this is the case, all the phases are so small that the resulting CP asymmetry is about 10  orders 
of magnitude smaller than needed.
%
%
However, the initial top quarks are bound in the bag, so light quarks can be propagating  at the energy much smaller than the top mass. The smallest possible scale is fixed by
  the weak interaction of quarks with the electroweak plasma outside the bag,
known as the screening mass $\sim g_wT$, which is few GeV. 
This is the natural scale to take: thus we will from now consider $E\sim m_b$ in the following. 
Another improvement one may try is to consider  locations of some points $outside$ the bag, selecting $r_{ij}$  as large as possible.
\ba
Prob_{t-\bar t}(r_1)&\sim&J\frac{m_b^4m_c^4 m_s^2}{64 E^5}\int dr_2 dr_3 2 r_2 r_3^2 f(r_1)^2 f(r_2) f(r_3).
\ea
Considering a radial bag of $N_W$ weak bosons having an exponential profile with the usual $W$ mass in the broken phase:
$
W(r)=\sqrt{\frac{N_w m_w^3}{\pi E_w}}e^{-m_w r},
$
we get that the probability of a top-minus-antitop quark escaping 
\be
\delta_{CP}=\tau J N_w^2 \frac{m_c}{16 E}\frac{m_b^4}{E^4}\frac{m_c^3}{m_w^3}\frac{m_s^2}{E_w^2}\sim 10^{-10}\left(\frac{N_w}{1000}\right)^2.
\label{best}
\ee
In the latter formula we made use of the lifetime of the bag denoted $\tau/m_W$, with $\tau\sim 6$ to bound the time integral over $x_1$. 

 The main lesson we got from this study
 is that the scales of both the quark energy $E$ and their traveling distances $r_{ij}$ in the loop amplitudes should be tuned individually,
 to maximize the effect. The main limitation come from the conditions of quark rescattering in  the plasma (the screaning masses)
 and the conditions that all phases $\delta_i$ should not be large,
 as well as the limitations coming from the $W$ field strength and correlation length.
 Another lesson is that in order to prevent cancellations between different flavors, one has to expand all the
 results till sign-definite answer is guaranteed.

The probability to find 3 antitops is actually proportionnal to $(1+\delta_{CP})^3\approx 1+3\delta_{CP}$,
 while it is $ (1-3\delta_{CP})$ for tops: it gives factor 3. Another factor
 3 appears because of the fact that each sphaleron event creates 3 units of baryon number, not one. 
 Together with baryon asymmetry (time integrated) sphaleron rates of $10^{-2}$ and $3*3*\delta_{CP}$ we arrive to our final estimate
 $
 \Delta B\sim 10^{-11\pm 1}
 $  
where one order of magnitude stands for our (perhaps optimistic) errors due to numerical factors ignored in the estimates. We conclude that
it is clearly in the same ballpark as the observed baryonic asymmetry of the Universe. Clearly, numerical factors can be detailed later, 
and the parameters of the cosmological model can be better tuned to get closer the right value.

Last but not least is the issue of the $sign$ of the asymmetry. Our formula (\ref{F})  has definite (positive) sign, that is to say more top quark escape the bag (note that the time direction is important, quarks are first created in the bag, then have more probability to escape). More antitops remain in the bags, with  more likely to be ``recycled" by the sphalerons: this produces more baryons than anti-baryons. Apparently we got the right sign for the baryon asymmetry.



{\bf Acknowledgements} I just was lucky to get invited and give a talk: all it contains
I learned with my collaborators over the years. I should also acknowledge very
useful recent discussion of these issues with J.Smit and A.Tranberg.

\end{document}